\documentclass[final]{svjour2}
\usepackage{graphicx}
\usepackage{rotating}
\usepackage{amssymb}
\usepackage{mathptmx}
\usepackage{xcolor}
\usepackage[numbers]{natbib}
\makeatletter
\bibpunct{}{}{,}{s}{}{,}

\begin{document}

\newcommand{\hdblarrow}{H\makebox[0.9ex][l]{$\downdownarrows$}-}
\title{Interplay between condensation energy, pseudogap and the specific heat of a Hubbard model in a n-pole approximation}

\author{A. C. Lausmann$^1$ \and E. J. Calegari$^1$ \and S. G. Magalhaes $^2$\and C. M. Chaves$^3$ \and A. Troper$^3$}

\institute{1: Laborat\'orio de Teoria da Mat\'eria Condensada, Universidade Federal de Santa Maria,\\ 97105-900, Santa Maria, RS, Brazil\\
%Department of Physics and Astronomy, Northwestern University,\\ Evanston, IL 60208, USA\\
%Tel.:\\ Fax:\\
\email{hanakadia@gmail.com}
\\2: Universidade Federal Fluminense, Av. Litor\^anea s/n, 24210-346 Niter\'oi, RJ, Brazil\\ 3: Centro Brasileiro de Pesquisas F\'{\i}sicas, Rua Xavier Sigaud 150, 22290-180, Rio de Janeiro, RJ, Brazil}

\date{XX.XX.2014}

\maketitle


\begin{abstract}

The condensation energy and the specific heat jump of a two-dimensional Hubbard model, suitable to discuss high-$T_c$ superconductors,
 is studied.
In this work, the Hubbard model is investigated  by  the Green's function method within a
 $n$-pole approximation, which allows to consider superconductivity with $d_{x^2-y^2}$-wave pairing.
In the present scenario, the pseudogap regime emerges when the antiferromagnetic (AF) correlations 
become sufficiently strong to move to lower energies the region around of the nodal point $(\pi,\pi)$ on the renormalized bands.
It is observed that above a given total occupation $n_T$, the specific heat jump $\Delta C$ and also the condensation energy 
$U(0)$ decrease signaling the presence of the pseudogap. 

%This same effect is noof the antiferromagnetic correlations on the condensation energy and on superconductivity are also investigated.

%PACS numbers: 74.70.Tx,74.25.Ha,75.20.Hr
\end{abstract}

\section{Introduction}

It is believed that the two-dimensional Hubbard model\cite{hubbard} is able to capture the essential physics of the 
high temperature superconductivity (HTSC) in copper-oxides\cite{Scalapino,Millis}. In such systems, understanding the interplay between 
the superconductivity and the pseudogap  regime could be the key to clarify the mechanisms behind the unconventional superconductivity.
Experimental results for some cuprates indicate a close relation among specific heat, condensation energy and the pseudogap\cite{loram,loram1,loram2,momono}. 
More precisely, due to the presence of a pseudogap on the normal state density of states, the jump in the specific heat and the superconducting condensation energy 
decrease below a given doping. Besides, according to references\cite{Millis,Millis1},
the HTSC phase diagram can be separated in two regimes: a weak coupling and
a strong coupling regime\cite{Millis,Millis1}. The weak coupling regime could be described, approximately, in terms of the 
conventional BCS superconductivity while the
strong coupling regime would be governed by unconventional superconductivity. In this scenario, the pseudogap is a property of 
the strong coupling regime\cite{Millis}. 
In this context, the investigation of the specific heat and the condensation energy of two-dimensional Hubbard model may give 
us important insights about 
the physics of the HTSC.

In the present work, the normal-state pseudogap and the superconducting regime of a two-dimensional Hubbard model is investigated within 
the Green's functions technique\cite{roth,edwards}. The pseudogap emerges on the strongly correlated regime in which the antiferromagnetic correlations associated with
the spin-spin correlation function 
$\langle \vec{S}_i\cdot\vec{S}_j\rangle$ becomes sufficiently strong to open a pseudogap in the region $(\pi,0)$ on the Fermi surface.
Such normal-state pseudogap is also observed in the $(\pi,0)$ point of the renormalized band.

\section{Model and method}

The repulsive ($U>0$) one band two-dimensional Hubbard model \cite{hubbard} studied here is
\begin{equation}
H=\sum_{\langle \langle ij \rangle\rangle \sigma} t_{ij}c_{i\sigma}^{\dag}c_{j\sigma} + \frac{U}{2}\sum_{i \sigma} n_{i,\sigma} n_{i,-\sigma}-\mu\sum_{i\sigma}n_{i\sigma}
\label{eqH1}
\end{equation}
which takes into account hopping to first and second nearest neighbors. The quantity $\mu$ represents the chemical potential,
 $n_{i,\sigma }=c_{i\sigma }^{\dag}c_{i\sigma }$ is the number operator and $c_{i\sigma }^{\dag }(c_{i\sigma })$ is the fermionic creation
(annihilation) operator at site $i$ with spin $\sigma =\{\uparrow ,\downarrow \}$.
We use the Green's function technique in the Zubarev's formalism\cite{zubarev}. The equation of motion of the Green's 
functions are treated within the $n$-pole approximation introduced 
by L. Roth \cite{roth,edwards}. In this procedure, a set of operators $\{\hat{A}_n\}$ is introduced in order to describe
 the most important excitations of the system.
 %of interest. 
 The $n$-pole approximation assumes that the commutator
$[\hat{A}_n,\hat{H}]$,  which appears in the equation of motion of the Green's 
functions, can be written as $[\hat{A}_n,\hat{H}]=\sum_m K_{nm}\hat{A}_m$
where the elements $K_{nm}$ are determined by anti-commuting both sides of this relation
 with the operator set $\{\hat{A}_n\}$ and taking the thermal average.
We get $\bf{K}=\bf{EN^{-1}}$
with
\begin{equation}
E_{nm}=\langle [[\hat{A}_n,\hat{H}],\hat{A}^{\dagger}_m]_+\rangle ~~~~~~\mbox{and}~~~~~~N_{nm}=\langle [\hat{A}_n,\hat{A}^{\dagger}_m]_+\rangle.
\label{eq4}
\end{equation}
In terms of $\bf{E}$ and $\bf{N}$, the Green's function matrix is 
%
%\begin{equation}
$\bf{G}(\omega)=\bf{N}(\omega\bf{N}-\bf{E})^{-1}\bf{N}$.
%\label{eq5}
%\end{equation}
%
Both $\bf{E}$ and $\bf{N}$ can be determined through equations  (\ref{eq4}) if the set of operators $\{\hat{A}_n\}$ is 
known. As we are interested 
in investigating both the normal and the superconducting regimes, we use the operator set\cite{edwards} $\{ \hat{A}_n\}=\{\hat{c}_{i\sigma},\hat{c}_{i\sigma}\hat{n}_{i-\sigma},\hat{c}_{i-\sigma}^{\dagger},\hat{n}_{i\sigma}
 \hat{c}_{i-\sigma}^{\dagger}\}$.
 %includes electron and hole operators\cite{edwards}.

%\subsection{Internal energy}

The energy per particle can be obtained from
%in terms of 
the Green's function following the procedure described by Kishore and Joshi \cite{kishore}.
For the Hubbard model introduced in equation (\ref{eqH1}), the internal energy per particle in the superconducting state is:
\begin{equation}
 E=\frac{1}{2L}\sum_{\vec{k},\sigma}\sum_{i=1}^{4}Z_{i,\vec{k}\sigma}
 (\varepsilon_{\vec{k}}+\mu+E_{i,\vec{k}\sigma})f(E_{i,\vec{k}\sigma}) -\mu n_T\, 
\label{eqE} 
\end{equation}
where $n_T=n_{-\sigma}+n_{\sigma}$ is the total occupation, $Z_{i,\vec{k}\sigma}$ are the spectral weights\cite{edwards} of the Green's function 
$G^{(11)}_{\vec{k},\sigma}=\langle\langle c_{\vec{k},\sigma};c_{\vec{k},\sigma}^\dagger \rangle\rangle$ and $f(\omega)$ is the Fermi function.
In the superconducting state, 
%The $E_{i,\vec{k}\sigma}$ are the superconducting state 
the renormalized bands are:
%defined as:
%
\begin{equation}
 E_{i,\vec{k}\sigma}=(-1)^{(i+1)}\sqrt{\omega_{j,\vec{k}\sigma}^{2}+
 \, \frac{(-1)^{(j+1)}|\gamma_{\vec{k}}|^{2}[(\varepsilon_{\vec k}+Un_{-\sigma}-\mu)^{2}-\omega_{j,\vec{k}\sigma}^{2}]}{n_{-\sigma}^{2}(1-n_{-\sigma})^{2}(\omega_{2,\vec{k}\sigma}^{2}-\omega_{1,\vec{k}\sigma}^{2})}}\, ,
\end{equation}
with $j=1$ if $i=1$ or $2$, and $j=2$ if $i=3$ or $4$, $\gamma_{\vec{k}}=2t\gamma(cos(k_xa)-cos(k_ya))$ is the gap function and
$\gamma$ is the superconducting order parameter with $d_{x^2-y^2}$-wave symmetry\cite{edwards}.
In the normal state, the renormalized bands are:
\begin{equation}
 \omega_{j,\sigma\vec k}=\frac{U+\varepsilon_{\vec k}+W_{\vec k,\sigma}-2\mu}{2}-(-1)^{(j+1)}\frac{X_{\vec k,\sigma}}{2}
\end{equation}
where $X_{\vec{k}}=\sqrt{(U-\varepsilon_{\vec{k}}+W_{\vec{k}\sigma})^{2} +4\langle n_{-\sigma}\rangle U(\varepsilon_{\vec{k}}-W_{\vec{k}\sigma})}$
and $\varepsilon_{\vec{k}}$ is the unperturbed band energy
%\begin{equation}
${\varepsilon }_{\vec{k}}=2t[\cos (k_{x}a) +\cos (
k_{y}a)] +4t_{2}\cos ( k_{x}a) \cos (k_{y}a)$ 
%\end{equation}
%
where $t$ is the first-neighbor and $t_{2}$ is the second-neighbor hopping
amplitudes.
% and $a$ is the lattice parameter. 
$W_{\vec k,\sigma}$, is a band shift that depends on the 
correlation function\cite{edwards,calegari} $\langle \vec{S}_i\cdot\vec{S}_j\rangle$.

The specific heat jump is $\Delta C=[\frac{(C_S-C_N)}{C_N}]_{_{T=T_c}}$ with $C_{S,N}=\frac{\partial E_{S,N}}{\partial T}$,
%in which
$E_S$ and $E_N$ being the energy per particle in the superconducting and in the normal sate, respectively.
%(given by equation (\ref{eqE})).
%and the normal state energy 
$E_N$ is
%also 
obtained from equation (\ref{eqE}) keeping the superconducting order 
parameter equal to zero $(\gamma=0)$.
% with the energy $E$ defined in equation \ref{eqE}. 
Now, let's define $U(T)=F_N-F_S$ as the difference between the normal ($F_N$) 
%Helmholtz free energy $F_N$
and the superconducting  ($F_S$) states Helmholtz free energy.
%with $F=E-ST$.
The superconducting condensation energy is defined as:
%giving by\cite{norman,loram3} $U(T=0)$, i. e.:
\begin{equation}
U(0)=E_N-E_S.
\label{eq6}
\end{equation}
%
%in which the $E_S$ is given by equation (\ref{eqE}).

\section{Results}

The main focus of the present work is the strong coupling regime in which unconventional superconductivity may occurs. For this purpose,
we analyzed the renormalized bands, the superconducting condensation energy and the specific heat jump as a function of the total
occupation $n_T$
%=n_{-\sigma}+n_{\sigma}$ 
and of the interaction  $U$.

\begin{figure}
\begin{center}
\includegraphics[angle=-90,
  width=1\linewidth,
  keepaspectratio]{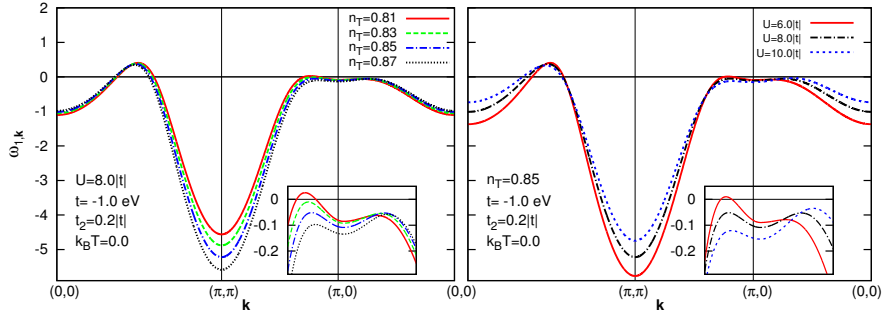}
\end{center}
\caption{(Color online) The left panel shows the renormalized band $\omega_{1,\sigma\vec k}$ for different occupations. In the inset, the region of the $(\pi,0)$ point shows the 
pseudogap for $n_T\gtrsim0.83$. In the right panel, the bands for $n_T=0.85$ and different interactions $U$. In the inset, the region of the $(\pi,0)$ point shows the 
pseudogap for $U\gtrsim 8.0|t|$.}
\label{fig1}
\end{figure}
\begin{figure}
\begin{center}
\includegraphics[angle=-90,
  width=0.6\linewidth,
  keepaspectratio]{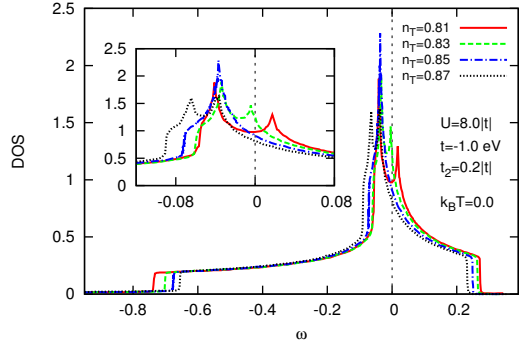}
\end{center}
\caption{(Color online) The density of states (DOS) for different occupations $n_T$. The vertical line in $\omega=0$ indicates the position of the Fermi energy $\epsilon_{F}$.
The inset shows the region near the Fermi energy $\epsilon_{F}$. For $n_T\gtrsim 0.83$ the DOS at $\epsilon_{F}$ is decreased indicating the presence of a pseudogap.
}
\label{fig1.1}
\end{figure}

Figure \ref{fig1} shows the renormalized band $\omega_{1,\sigma\vec k}$. In  the left panel, 
%\textcolor{red}{renormalized band 
$\omega_{1,\sigma\vec k}$ is shown for different values of total the occupation $n_T$. 
The inset displays the region near the point $(\pi,0)$ in which a pseudogap develops when the occupation is increased. For instance, when $n_T=0.81$, 
the band intersects the Fermi energy $\epsilon_F$, but, 
for $n_T=0.85$ the band does not reaches the Fermi energy giving rise to a pseudogap between the band and $\epsilon_F$.
The right panel shows $\omega_{1,\sigma\vec k}$ for $n_T=0.85$ and different 
interactions $U$. The  inset highlights a pseudogap on $(\pi,0)$ for $U=8.0|t|$ and $U=10.0|t|$ and the absence of pseudogap for $U=6.0|t|$. 
In the $n$-pole approximation used in this work\cite{roth,edwards} the Green's functions naturally present a pole structure which contains
the spin-spin correlation function\cite{edwards} $\langle \vec{S}_i\cdot\vec{S}_j\rangle$.
In the present scenario the pseudogap emerges when the correlation function $\langle \vec{S}_i\cdot\vec{S}_j\rangle$ becomes sufficiently
strong to move to lower energies the region of the nodal point $(\pi,\pi)$ of the renormalized band $\omega_{1,\sigma\vec k}$. 
This occurs because the renormalized band $\omega_{1,\sigma\vec k}$ is deeply influenced by the momentum structure of the spin-spin correlation
function\cite{Eleonir2011} $\langle \vec{S}_i\cdot\vec{S}_j\rangle$. Due to the antiferromagnetic character\cite{edwards} of $\langle \vec{S}_i\cdot\vec{S}_j\rangle$, 
the region of the nodal point $(\pi,\pi)$ of  $\omega_{1,\sigma\vec k}$  is strongly affected (see figure \ref{fig1}). As a consequence a pseudogap arises at the anti-nodal point $(\pi,0)$.
Moreover, the $\langle \vec{S}_i\cdot\vec{S}_j\rangle$ is very sensitive to $n_T$ and $U$,
 indeed, $|\langle \vec{S}_i\cdot\vec{S}_j\rangle|$ increases\cite{nolting} with $n_T$ and $U$. Therefore, when $|\langle \vec{S}_i\cdot\vec{S}_j\rangle|$ 
reaches a critical value $|\langle \vec{S}_i\cdot\vec{S}_j\rangle|_c$ the pseudogap emerges.

 The density of states (DOS) for the renormalized band $\omega_{1,\sigma\vec k}$ is shown in figure \ref{fig1.1} for different occupations $n_{T}$.
The vertical line in $\omega=0$ indicates the position of the Fermi energy $\epsilon_{F}$ and the model parameters are shown in the figure.
%Notice that 
When $n_T$ increases, the correlations become stronger, resulting in a narrowing of the density of states. However, the most important feature
observed in the DOS is the reduction of the DOS on $\epsilon_{F}$ for $n_T\gtrsim 0.83$. Such reduction is an effect of the presence of a pseudogap
in the strong correlated regime of the system.

\begin{figure}
\begin{center}
\includegraphics[angle=-90,
  width=1.0\linewidth,
  keepaspectratio]{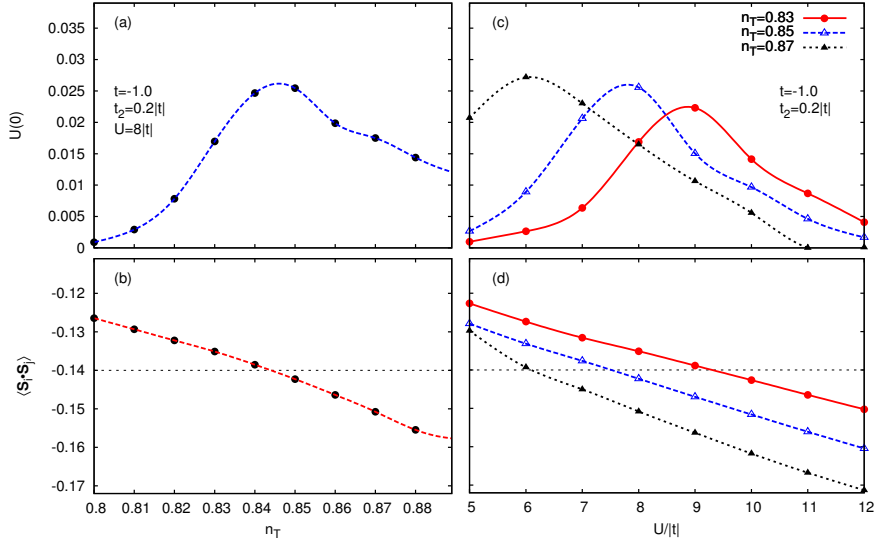}
\end{center}
\caption{(Color online) In (a), the condensation energy as a function of the total occupation $n_T$. (b) The 
behavior of the spin-spin correlation function for the same model parameters considered in (a). In (c) the condensation energy as a function of the interaction $U$ for different
occupations. (d) The spin-spin correlation function as a function of $U$ for the same model parameters considered in (c). Details about these results are given in the text.}
\label{fig2}
\end{figure}

The condensation energy $U(0)$ as a function of the total occupation $n_T$ is shown in figure \ref{fig2}(a).
Notice that $U(0)$ increases with $n_T$, reaches a maximum and then start to decrease.
The depletion of $U(0)$ for a $n_T$ greater than a given value is related to the development of a pseudogap near the anti-nodal points on the Fermi surface. 
This result is in qualitative agreement with experimental data obtained for some cuprate systems \cite{loram1,loram2}. 
%The pseudogap, can be also observed on $(\pi,0)$ point of quasiparticles bandas shown in the left panel of figure \ref{fig1}.
Figure  \ref{fig2}(b) shows 
%the enhancement of
$\langle \vec{S}_i\cdot\vec{S}_j\rangle$ as a function of $n_T$. The horizontal dotted line
indicates approximately the value of $\langle \vec{S}_i\cdot\vec{S}_j\rangle$, from which the system enters in the underdoped strong coupling regime.
Figure \ref{fig2}(c) displays the behavior of the condensation energy $U(0)$ as a function of the interaction $U$ for several occupations.
It is interesting to note that there is an optimal value of $U$ which produces a maximum $U(0)$. However, such optimum value changes with the occupation $n_T$.
This feature is associated to the opening of the pseudogap which occurs in the strong coupling regime. Therefore, if $n_T$ decreases, 
a higher value of $U$ is necessary for the system to access the strong coupling regime. The correlation function $\langle \vec{S}_i\cdot\vec{S}_j\rangle$ shown
in figure \ref{fig2}(d) may serve as a parameter to indicate that the system is reaching the strong coupling regime.
As in figure \ref{fig2}(b), the horizontal dotted line indicates approximately the value of $\langle \vec{S}_i\cdot\vec{S}_j\rangle$,
 from which the system enters in the underdoped strong coupling regime.
 The results for condensation energy $U(0)$ versus $n_{T}$, figure \ref{fig2}(a), are in qualitatively agreement with a method bansed on the resonanting valence bond (RVB) spin liquid \cite{leBlanc}
 and also with results from the fluctuation-exchange (FLEX) approximation \cite{Yanase}. In the FLEX approximation, $U(0)$ increases with  $U$ but does not present a maximum like the one observed in figure \ref{fig2}(c).
 There are no available results for  $U(0)$ versus $U$ in the RVB method \cite{leBlanc}.

\begin{figure}
\begin{center}
\includegraphics[angle=-90,
  width=0.6\linewidth,
  keepaspectratio]{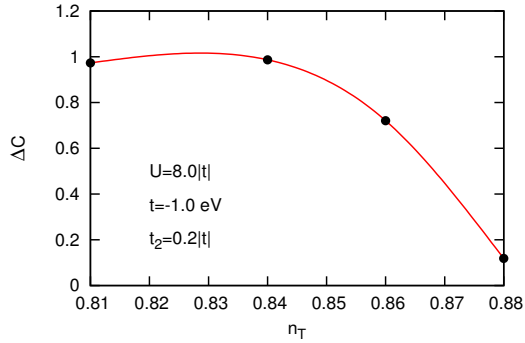}
\end{center}
\caption{(Color online) The jump in the specific heat as a function of the total occupation (see the text).}
%\textcolor{red}{Details about this result are given in the text.}}
\label{fig4}
\end{figure}

The figure \ref{fig4} shows the specific heat jump $\Delta C$ as a function of $n_T$. 
Notice that initially the $\Delta C$ increases slightly with $n_T$ but, above $n_T\approx 0.83$, $\Delta C$ starts to decrease.
The decreasing in $\Delta C$ is an evidence of the presence of a pseudogap in the underdoped regime and is close related to the development 
of a pseudogap on the density of  states (DOS) (see figure \ref{fig1.1}). The decreasing of the  DOS on $\epsilon_F$ when $n_T$ increases, is directly related
to the pseudogap on $\epsilon_F$ and agrees with a high-resolution photoemission study \cite{ino} of La$_{2-x}$Sr$_x$CuO$_4$ which suggests that a pseudogap is the main
responsible for the similar behavior between the specific heat jump and the DOS($\epsilon_F$) observed in the underdoped regime.
This result for $\Delta C$ agrees at least qualitatively, with a method based on the resonating valence bond (RVB) spin liquid \cite{leBlanc}
in which the specific heat jump and the condensation energy decrease due to the opening of a pseudogap in the underdoped regime.
%It is important to note that above a given value of $n_T$
%the $\Delta C$ starts to decreases. The decreasing in $\Delta C$ is an evidence of the presence of a pseudogap in the underdoped regime. 
Also, the result  for $\Delta C$ shown in figure \ref{fig4} is in qualitative agreement with experimental data for some cuprates \cite{loram,loram1}.

\section{Conclusions}

In this work we have investigated the superconducting condensation energy $U(0)$ and the specific heat jump $\Delta C$ of a two-dimensional Hubbard model.
The results show that both $U(0)$ and $\Delta C$ decreases in the strong coupling underdoped regime. It has been verified that this behavior
is related to the opening of a pseudogap at the anti-nodal point $(\pi,0)$ on the renormalized band $\omega_{1,\sigma\vec k}$. In the strong coupling regime,
the correlation function $\langle \vec{S}_i\cdot\vec{S}_j\rangle$ presents in the band shift becomes sufficiently strong
to move to lower energies the renormalized band $\omega_{1,\sigma\vec k}$ in the region of the nodal point $(\pi,\pi)$ and 
as a consequence, a pseudogap opens in the $(\pi,0)$ point. The results obtained here corroborate the scenario that attributes the pseudogap to the strong 
correlations present in the underdoped regime\cite{Millis1}.

\begin{acknowledgements}
This work was partially supported by the Brazilian agencies CNPq,
%(Conselho Nacional de Desenvolvimento Cient\'{\i}fico e Tecnol\'ogico),
CAPES
%(Coordena\c{c}\~ao de 
%Aperfei\c{c}oamento de Pessoal de N\'{\i}vel Superior), 
and FAPERGS.
%(Funda\c{c}\~ao de Amparo \`a Pesquisa do Rio Grande do Sul) 
%and FAPERJ.

\end{acknowledgements}

%\pagebreak

\end{document}